\documentclass[12pt,a4paper]{article} 

\usepackage{graphicx} 

\usepackage{longtable}

\usepackage{setspace}

\usepackage[paper=a4paper,left=25mm,right=25mm,top=25mm,bottom=25mm]{geometry}

\usepackage{booktabs}

\usepackage[footnotesize]{caption}

\usepackage{picins}

\usepackage{pifont}

\usepackage{booktabs}

\usepackage{colortbl}

\definecolor{hellgrau}{rgb}{0.95,0.95,0.95}

\usepackage{subfig}

\usepackage{listings}

\begin{document}


\thispagestyle{empty}

\vspace*{1cm}

\begin{center}

 \section*{\bf Analysing tax evasion dynamics via the Ising model}

\vspace{0.5cm}

\renewcommand{\thefootnote}{\fnsymbol{footnote}}

Georg Zaklan$^{*}$\\

Frank Westerhoff\footnote[1]{Department of Economics, University of Bamberg, Feldkirchenstrasse 21, D-96045 Bamberg, Euroland.} \\

Dietrich Stauffer\footnote[7]{Institute of Theoretical Physics, University of Cologne, Zuelpicher Strasse 77, D-50937 Koeln, Euroland.\vspace{0.4cm}}\\[0.3cm]

\vspace{0.5cm}

Comments welcome. Contact: georg.zaklan@uni-bamberg.de\\[1cm]

\vspace{0.5cm}

\today 

\end{center}

\vspace{0.55cm} 

  \begin{abstract}

\noindent 

We develop a model of tax evasion based on the Ising model. We augment the model using an appropriate enforcement mechanism that may allow policy makers to curb tax evasion. 



With a certain probability tax evaders are subject to an audit. If they get caught they behave honestly for a certain number of periods. Simulating the model for a range of parameter combinations, we show that tax evasion may be controlled effectively by using punishment as an enforcement mechanism.
\\[0.3cm]

[Keywords: Opinion dynamics, Sociophysics, Ising model.]\\[0.0cm]



\end{abstract}

\newpage

\renewcommand{\thefootnote}{\arabic{footnote}}

\section{Introduction}

Despite significant progress in expanding Allingham and Sandmo's (1972) seminal theoretical paper to compensate for its weakness in predicting sufficiently high levels of individual tax compliance, the way enforcement mechanisms work from a dynamic perspective still needs further attention. Specifically, a new line of research is pursuing a multi-agent based simulation approach (MABS), to analyse how different enforcement measures influence the compliance behaviour of individuals in complex systems, where possibly different agents interact. A literature review of MABS-models regarding income tax evasion can be found in Bloomquist (2006).

We find it especially worthwhile to investigate Allingham and Sandmo's theoretical prediction that both an increase in audit probability and an increase in tax penalty enhance tax compliance. We perform our analysis in the light of group influence on individual behaviour and find that both measures work to reduce tax evasion. Our study is related to two other MABS-studies: Davis, Hecht and Perkins (2003) (DH\&P) and Bloomquist (2004), who develops a powerful tax compliance simulator (TCS), adressing various possible triggers of tax evasion.  Like DH\&P we assume that tax payers who are audited become honest upon audit. While Bloomquist assumes that an individual becomes more risk averse when seeing someone in her network being audited, we equip caught tax evaders with a memory. It reminds them to remain honest for a certain number of time periods after non-compliance was detected. We interpret the enforced period of honesty as a punishment. The severity of punishment obviously depends on the number of periods detected tax cheaters need to remain honest for.

The importance we accord to group influence regarding an individual's decision whether to evade or not mainly stems from the work of DH\&P, who stress that viewing others evade a previously honest tax payer becomes susceptible to evasion with some probability. The study of Korobow et al. (2007) also illustrates that especially group effects are important for an individual's decision whether to evade or not. They find that the existence of social networks diminishes compliance.

To further analyse how tax compliance develops over time when individual decision making is subject to group influence we use the Ising model, a simple model from physics describing how particles interact under different temperature levels. We find this modeling framework particularly appropriate, because it attaches a large probability to a state in which an individual takes on the type that dominates her neigbourhood. To conform with the Ising model, we assume that only two types of individuals exist, honest citizens and tax evaders.

\section{The model}

We use a 1000 $\times$ 1000 grid square lattice, where in every time period each lattice site is inhabitated by an individual (=spin $S_i$) who can either be an honest tax payer $S_i = +1$ or a cheater $S_i = -1$ , trying to at least partially escape her tax duty. It is assumed that initially everybody is honest. Each period individuals can rethink their behaviour and have the opportunity to become the opposite type of agent  they were in the previous period.

The neighbourhood of every individual is composed of four people, agents  to the north, west, east and south. Each agent's social network may either prefer tax evasion or reject it. Various degrees 

of homogeneity regarding either position are possible. An extremely homogenous group is entirely  made up of honest people or of evaders. No majority regarding either position exists only when the neighbourhood is completely mixed up. This is the case when two individuals repectively prefer each position.

Individual decision making depends on two factors. On the one hand, the type of network every agent is connected with exerts influence on what type of citizen she becomes in the respective period. On the other hand, peoples' decisions are partly autonomous, i.e. they are not influenced by the constitution of their vicinity. The autonomous part of individual 

decision making is responsible for the emergence of the tax evasion problem, because some initially honest tax payers decide to evade taxes and then exert influence on others to do so as well. How large the influence from the neighbourhood is can be controlled through the ``social temperature'' parameter, $T$ (units: $J/k_B$).

Total energy is given by the Hamiltonian $H=-\sum_{<i,j>}J_{ij} S_i S_j-B\sum_{i}S_i$. The sum runs over all nearest neighbour pairs of spins, $S_i$ and $S_j$. $J_{ij}$ is the coupling between spins, which we assume to be constant ($J_{ij}=J$) for all neighbouring spins. B is a positive parameter. It denotes the importance of the magnetic field for 

the total energy; in the tax evasion context it could be interpreted as the influence of mass media, but we do not use it here (i.e. $B=0$). The magnetisation  is given by summing the corresponding values of all spins ($=\sum_{i}S_i$).

$I_e=S_i \cdot \left(\sum_j S_j\right)$, multiplied by $-J$, denotes the energy resulting from the interaction between the considered individual and her four closest neighbours. It is calculated by adding up the products of the respective individual's type (spin $S_i$) and the type of each of her four neighbours (spins $S_j$). It is known since decades that for $T > T_c = 2/\mbox{ln}(1 + \sqrt 2) \approx 2.2$ half of the spins are +1 and the other half $-1$, while for $T < T_c$ there is a majority for one direction.

We simulate the Ising model by performing a spin-flip only if a random number between 0 and 1 is smaller than the normalised probability ($p$) of a spin-flip:  $p=\mbox{exp}(-\Delta E/k_B T)/(1+\mbox{exp}(-\Delta E/k_B T)$.

The following table illustrates the probabilities of a spin-flip, given 

a range of possible structures for the neighbourhood and the different temperature levels we used in our simulations (cp. figures 1 to 4).

\vspace{0.5cm}

\begin{center}

\begin{tabular}[h]{l|c|c|c|c|c}

\cline{2-6} 

\multicolumn{1}{c}{} & \multicolumn{5}{|c}{Temperature T} \\ 

\cline{2-6} 

    &   $T=0.25$ &    $T=2.0$ &   $T=2.5$ &   $T=3.0$ &   $T=25$ \\

\hline

$I_e$ & \multicolumn{5}{c}{\textbf{Probability of a spin-flip}} \\

\toprule

$I_e=-4$ & $\approx$ 1  &  0.982014 &  0.960835 &  0.935031 &  0.579325 \\

$I_e=-2$ & $\approx$ 1 &  0.880797 &  0.832019 &  0.791392 &  0.539915 \\

$I_e=0$ & 0.5 & 0.5 & 0.5 & 0.5 & 0.5 \\

$I_e=2$ &  $\approx$ 0 &  0.119203 &   0.1679815 &  0.208608 &  0.460085 \\

$I_e=4$ & $\approx$ 0 &  0.017986 &  0.0391655 &  0.064969 &  0.420676\\

\toprule

\end{tabular}

\end{center}

\vspace{0.5cm}

\noindent The higher the level of the temperature is, the 

more a probability near $p=1/2$ is accorded to a spin-flip, regardless of whether it implies a reduction ($I_e=-4$ and $I_e=-2$) or an increase in energy ($I_e=4$ and $I_e=2$). 

On the other hand, the lower the temperature is (e.g. $T=0.25$), the more certain it becomes that flips, which cause energy to fall, take place and   

that flips, which cause energy to rise, do not occur. 

Finally, in the intermediate case where energy remains unchanged by a spin-flip ($I_e=0$), the probability of a flip is equal to $p=1/2$, regardless of the temperature level.

Applied to tax evasion we can interpret the model as follows: Tax evaders have the greatest influence to turn honest citizens into tax evaders if they constitute a majority in the respective neighbourhood. 

If the majority evades, one is likely to also evade. On the other hand, if most people in the vicinity are honest, the respective individual is likely to become a tax payer if she was a tax evader before. For very low temperatures, for instance $T=0.25$, the autonomous part of decision making almost completely disappears. Individuals then base their decisions solely on what most of their neighbours do. A rising temperature has the opposite effect. The individuals then decide more autonomously.

We further introduce a probability of an efficient audit ($p_a$). If tax evasion is detected, the individual must remain honest for a number of periods to be specified. We vary this amount of time and denote the period of time that the caught tax evaders are punished for by the variable $k$.\footnote{By adding the number of tax evaders and honest citizens together in one time period, taking into account the positive or negative sign every agent is marked with, one can also easily find the magnetisation for each time period.} One time unit is one sweep through the entire lattice. Appendix A contains a Fortran source code for our model. 

For background information on the Ising model and how to simulate it, see Stauffer et al. (1988).

Already F\"ollmer (1974) applied the Ising model to economics.

\section{Dynamics of the model}

We simulate tax evasion dynamics for various temperature levels and differently severe punishments ($k$). When $k$ equals zero, no punishment is present and the model describes the baseline Ising model. Various degrees of punishment are introduced for different temperatures, by setting $k$ consecutively equal to 10 and 50 periods for all considered levels of the temperature. The probability of an audit is sequentially increased in steps of one percent, from 0 to 100 percent. For a given probability of an audit the dynamics of tax evasion (measured as portion of the entire population) is depicted over the range of 300 time steps. Our three-dimensional illustrations thus depict 101 single time series, one for each possible level of the probability of an audit. Additionally we use two-dimensional illustrations to better convey the form of the tax evasion dynamics, when the probability of an audit is either at a realistic level ($p_a=0.05$) or at a rather high level ($p_a=0.9$).

\vspace{0.3cm}

\begin{center}

---$\quad$Figure $1$ goes about here$\quad$---

\end{center}

\vspace{0.3cm}

\noindent In figure 1 we set $T=25$. If the penalty is high enough (e.g. $k=50$), tax evasion can be reduced to $0$\% in the short run, given that the probability of an audit is sufficiently high. In the case of a penalty duration of $50$ periods and a probability of an audit of $90$\% (left panel in the second row of figure $1$), within only a few periods each individual eventually is compelled to remain honest. This happens, because spins flip relatively often at this temperature, which is far above the critical level. The peaks we observe in the level of the tax evasion, result from the fact that $90$ percent of the initial large number of tax evaders gets caught and after $k$ iterations simultaneously becomes free to decide whether to evade or not. Roughly half of them choose to become non-compliant again after being regiven the opportunity to evade. Moreover consecutive peaks in non-compliance diminish less over time and it takes longer until perfect compliance is established the further out on the time scale the evolution of tax evasion is considered at. When allowing for more time to pass, one can see that evasion eventually does not hit the mark of zero percent any more. After around $8000$ time steps an equilibrium level of about $2$\% non-compliance is attained, because the number of agents who can freely decide which type to take on stabilises at a level consistent with this portion of tax evasion.

If punishment is set equal to $10$ periods (right panel in the same row) tax evasion only approaches zero percent and finally comes to rest at a level of $9$\%. 

Because the length of punishment now is too small to reach full compliance, the peaks are somewhat greater than at $k=50$: Additionally to those individuals who are released after $k$ periods from having to remain honest there now also exist other individuals who may be tax evaders as well.\footnote{Some of the tax evaders who were not punished, because the duration of the punishment is too short to establish full compliance, also remain tax evaders in the subsequent periods.}\\[-0.25cm]

\noindent In the two corresponding time series plots where $k$ is equal to either $50$ or $10$ periods, but the probability of an audit is much lower ($p_a=0.05$), one can describe the dynamics similarly as above. Obviously the probability of an audit now is too low to reach full compliance, even if $k=50$ (cp. row three of figure 1). These two pictures illustrate well that punishment is a suitable enforcement mechanism when the probability of an audit is set to a realistic level. The more periods individuals are forced to remain honest in the case of detection, the lower the resulting equilibrium level of non-compliance apparently is. For $k=50$ it is equal to $21$\% and for $k=10$ the equilibrium level of tax evasion amounts to $39$\%. \\[-0.25cm]

\vspace{0.0cm}

\begin{center}

---$\quad$Figures $2$ and $3$ go about here$\quad$---

\end{center}

\vspace{0.3cm}

\noindent We also consider two other temperatures above $T_c$, $T=3$ and $T=2.5$. The lower the temperature is, the slower adjustment towards the equilibrium in the baseline model occurs (i.e. $50$\% non-compliance).\footnote{The amount of time necessary for this equilibrium of the Ising model to be reached under different temperatures can be read off in either of the three-dimensional illustrations in the figures $1$ to $3$, when considering the evolution of the tax evasion over time, given that $p$ is equal to $0$\%.} As individuals become tax evaders more slowly the tax evasion problem is less pronounced already from the beginning compared to higher temperatures. But, because spins flip less frequently at lower temperatures, the same enforcement mechanisms may work less efficiently in the short run than at higher temperatures. When considering either of the time series with $k=50$ and $p_a=0.9$ (at $T=3$ or $T=2.5$) one clearly sees that evasion cannot be reduced to zero percent any more. On the other hand, if the temperature is at $25$ everybody becomes an evader within only a few periods, so that the enforcement mechanism quickly entails the entire population, given that the probability of an audit is sufficiently high. Yet, for the considered low temperatures $T=3$ and $T=2$, the enforcement mechanism does not encompass every person any more, because it takes longer that all individuals once take on the type of a tax evader.

Thus full compliance cannot be established any more in the short run. On the other hand, when looking at the long-run, one can see that the equilibrium levels of tax evasion are the lower the smaller social temperature is.

\vspace{0.3cm}

\begin{center}

---$\quad$Figure $4$ goes about here$\quad$---

\end{center}

\vspace{0.3cm}

\noindent Finally, for the sake of completeness, we also introduce a temperature level below $T_c$ ($T=2$). At such low temperatures individuals seldomly decide to become non-compliant, because their vicinity which is mostly compliant on average exerts strong influence on them to be honest as well. Therefore the equilibrium levels of tax evasion are smaller than at higher temperatures. Obviously, also for this temperature a higher degree of enforcement works to reduce non-compliance more.

\section{Conclusion}

Considering that individuals are likely to be influenced in their decision to evade taxes by their immediate neighbours, we found that regardless of how strong group influence may be, enforcement always works to enhance tax compliance. Both, a higher probability of an audit and a larger punishment work together to enhance tax compliance. To exhaust the model's explanatory power regarding how group influence affects overall compliance, it appears interesting to perform a similar analysis for different initialisations (e.g. everybody is dishonest in the beginning). Also, it seems interesting to consider if and under what circumstances the system may show large fluctuations in tax compliance and whether these effects can be controlled, with the aim of fixing compliance at a high level.

We would like to thank Martin Hohnisch and Sabine Pittnauer for their valuable comments and their time for fruitful discussions.

\newpage

\section*{\hspace*{-2ex}Appendix A: Fortran  source code for a single time series}

\setstretch {0.969}

\small

\lstset{language=Fortran}

\begin{lstlisting}[numbers=left,breaklines]
!Input data: L=1000,k=50,T=25,mcstep=300,p=0.9
program tax_evasion
implicit none 
!Declaration of parameters and variables
integer,parameter :: L=1000
integer,parameter :: both=L*L
integer,parameter :: Lmax=(L+2)*L
integer,parameter :: Lp1=L+1
integer,parameter :: L2pL=L*L+L
integer,parameter :: L2p1=2*L+1
real,parameter :: T=25.0
integer,parameter :: mcstep=300
integer,dimension(L*(L+2)) :: is
real,dimension(-4:4) :: iex
integer,dimension(L*(L+2)) :: mem 
integer :: ie,m,mc,hon,i,j,ev,k,ibm=1    
real :: ex,p,ip
!Probability of an audit (p) is set to 90%
p=0.9
ip=(2*p-1.0)*2147483648.0
!Number of periods tax evaders need to remain honest if audited (k)
k=50  
!Initialisation: Set everybody to honest and their memory to zero
do m=1,Lmax
mem(m)=0
is(m)=1
end do
!Spin-flip probabilities
do ie=-4,4,2
ex=exp(-ie*2.0/T)
iex(ie)=(2.0*ex/(1.0+ex)-1.0)*2147483648.0
ibm=ibm*65539
end do
write(*,*) p,0,0
!Dynamics of tax evasion is simulated over  mcstep+1 time steps
do mc=1,mcstep
!Set counter of honest individuals to zero
hon=0
do i=Lp1,L2pL		    
!First periodic border constraint
if(i.eq.L2p1) then 
do j=1,L
is(j+L2pL)=is(j+L)
end do
end if
!Audited tax evaders must remain honest for k periods
if(mem(i).gt.0) then
mem(i)=mem(i)-1
is(i)=1
else
ie=is(i)*(is(i-1)+is(i+1)+is(i-L)+is(i+L))
ibm=ibm*16807
if(ibm.lt.iex(ie)) is(i)=-is(i)
end if
!Counting the number of honest citizens 
if (is(i).eq.1) hon=hon+1
ibm=ibm*16807
!Each audited tax payer obtains a memory
if(ibm.lt.ip.and.is(i).eq.-1) mem(i)=k
end do 
!Second periodic border constraint
do j=1,L
is(j)=is(j+both)    
end do
!The number of tax evaders 
ev=both-hon
!For the time steps 1 to 300 print the same quantities as for mc=0
write(*,*) p,mc,ev/real(both)
end do !End time step (mc) loop
end program tax_evasion
\end{lstlisting}


\newpage

\setstretch {1.5}

\newpage

\pagestyle{empty}



\setstretch {1}


\begin{center} 

\textbf{Temperature: 25}\\[0.8cm] 

\begin{tabular}{c}

\hspace{-1.9cm}\textbf{$\;\;\;\qquad\qquad k=50$} \hspace{6.5cm} \textbf{$\;\;k=10$}\\[0.0cm]

\includegraphics[width=7.9cm]{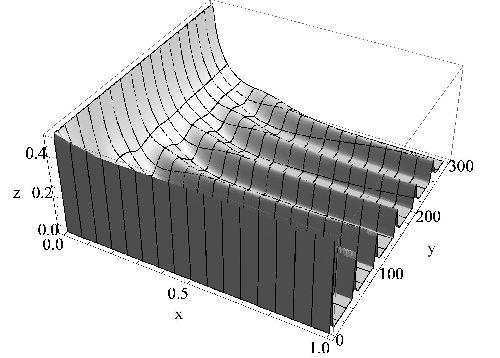}\hspace{-0.0cm}

\includegraphics[width=8.3cm]{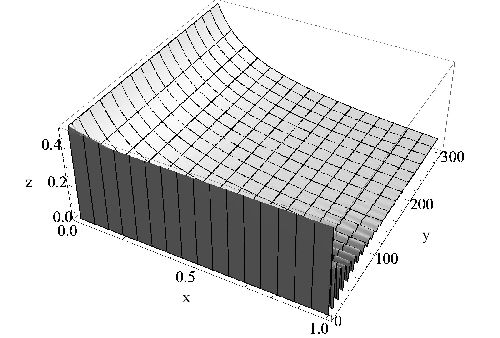} \\[0.8cm]

\noindent \hspace{0.0cm}\textbf{$\;\;\;\;k=50$, $p_a=0.9$} \hspace{3.9cm}  \textbf{$\;\;\;\;\;\;\;\; k=10$, $p_a=0.9$}\\[0.0cm]

\hspace{-0.95cm}

\includegraphics[width=7.0cm]{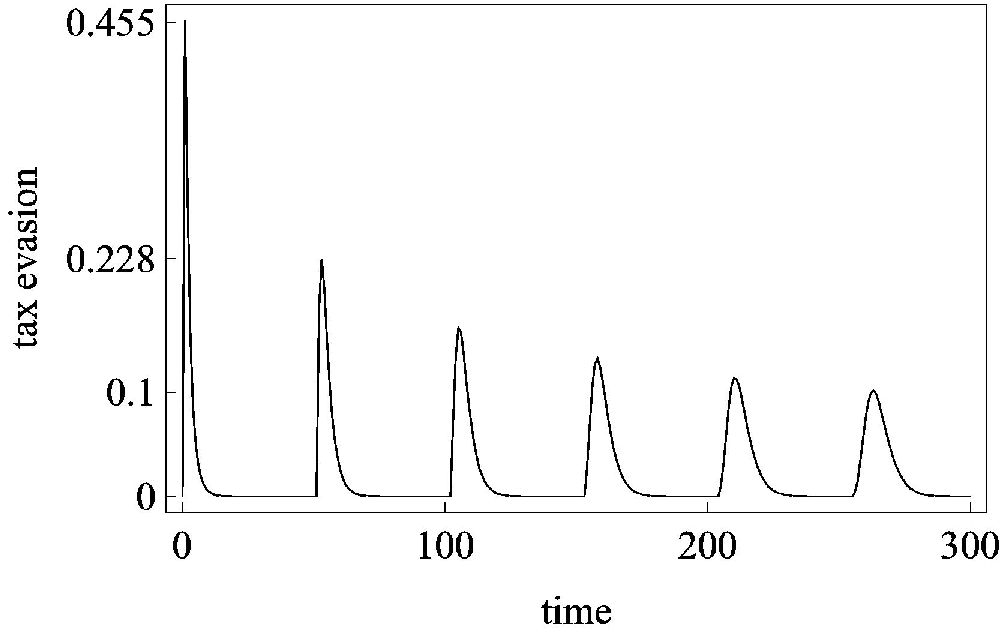}

\hspace{0.8cm} 

\includegraphics[width=7.0cm]{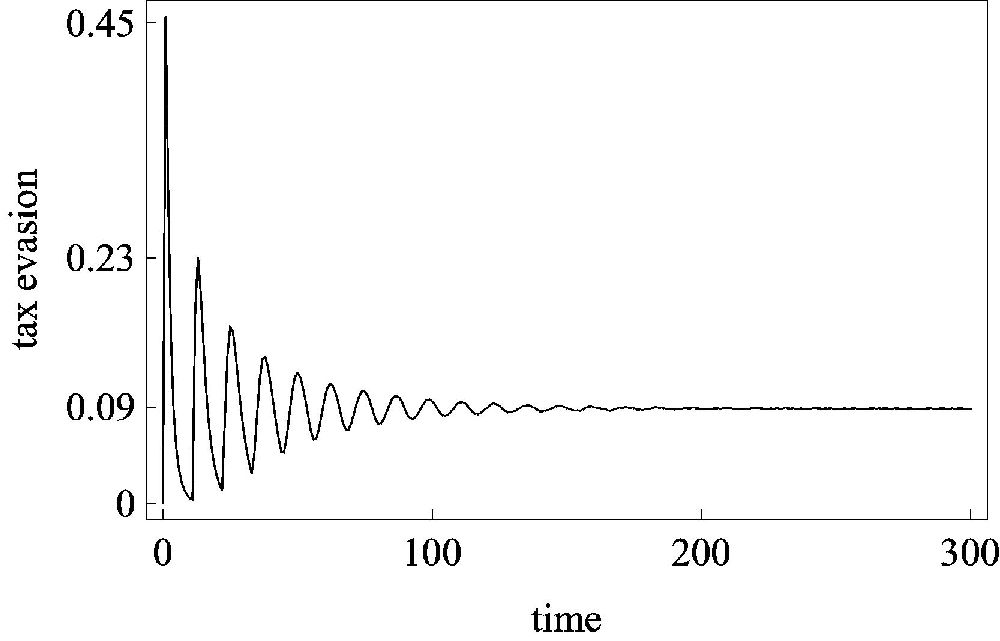} \\[0.8cm]

\noindent \hspace{0.0cm}\textbf{$\;\;\;\;k=50$, $p_a=0.05$} \hspace{3.9cm}  \textbf{$\;\;\;\;\;\;\;\; k=10$, $p_a=0.05$}\\[0.0cm]

\hspace{-0.95cm}

\includegraphics[width=7.0cm]{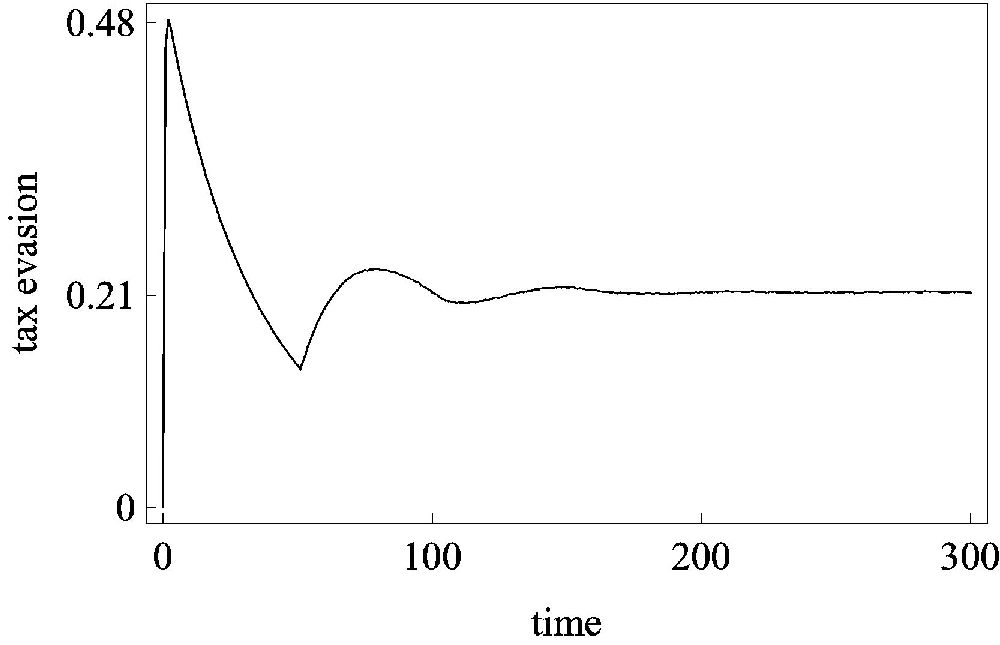}\hspace{0.8cm} 

\includegraphics[width=7.0cm]{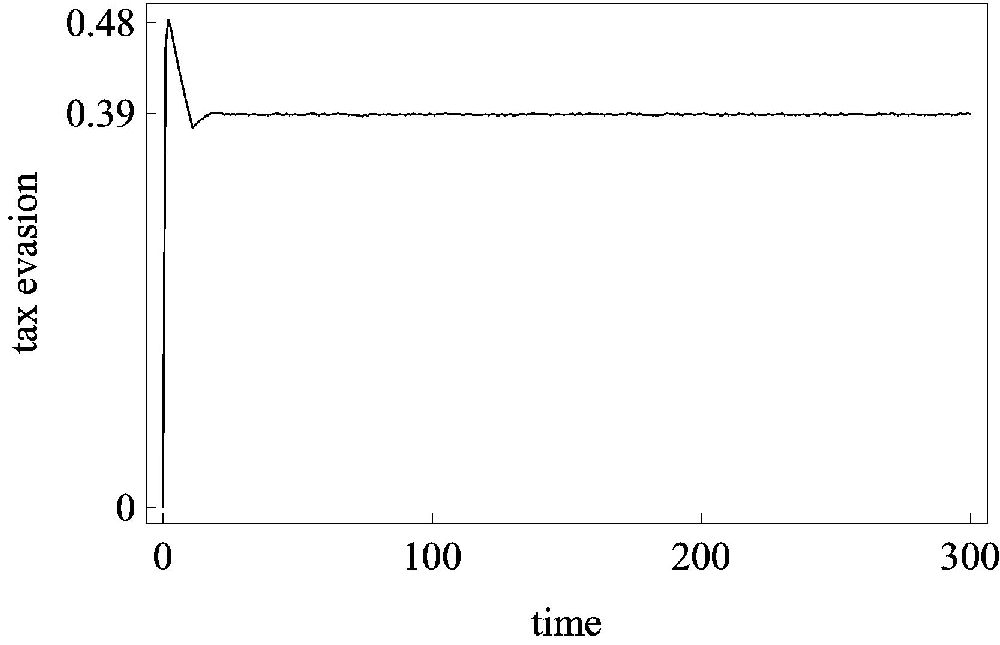}

\end{tabular}

\end{center}

\vspace{0.6cm}

{\footnotesize{\noindent Figure 1: In the first two pictures, the x, y and z-axis denote the probability of an audit $p_a$ (x), the considered time period (y) and the corresponding tax evasion portion (z), respectively.

These two pictures thus show the tax evasion dynamics when holding the social temperature constant at 25 and 

when controlling for the number of periods $(k)$ that a detected tax evader must remain honest for. If either $k$ or $p_a$ is equal to zero we get the standard Ising model. When $k$ and $p_a$ are nonzero and positive, the augmented version of the Ising model, i.e. our tax evasion model, applies.

The remaining four pictures visualise specific time series, by holding additionally the probability of an audit at a constant level. We leave the length of punishment unchanged at either 50 or 10 periods and depict the tax evasion dynamics for two different probabilites of an audit, i.e. $p_a=0.05$ and $p_a=0.9$.}}


\newpage

\vspace*{0.0cm}

\begin{center} 

\textbf{Temperature: 3.0}\\[1.0cm]  

\begin{tabular}{c}

\hspace{-1.9cm}\textbf{$\;\;\;\;\;\qquad\qquad k=50$} \hspace{6.5cm} \textbf{$k=10$}\\[0.0cm]

\includegraphics[width=7.8cm]{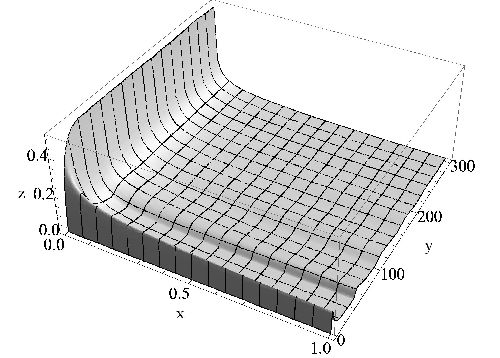}\hspace{-0.0cm}

\includegraphics[width=7.9cm]{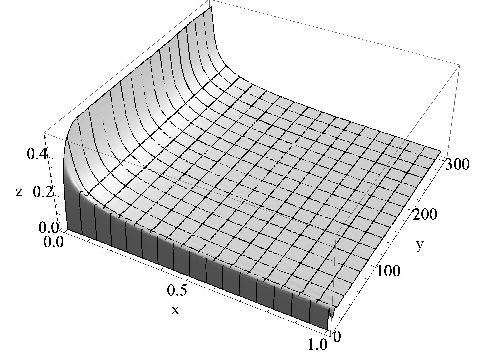} \\[0.8cm]

\noindent \hspace{0.0cm}\textbf{$\;\;\;\;k=50$, $p_a=0.9$} \hspace{3.9cm}  \textbf{$\;\;\;\;\;\;\;\; k=10$, $p_a=0.9$}\\[0.0cm]

\hspace{-0.4cm}

\includegraphics[width=7.0cm]{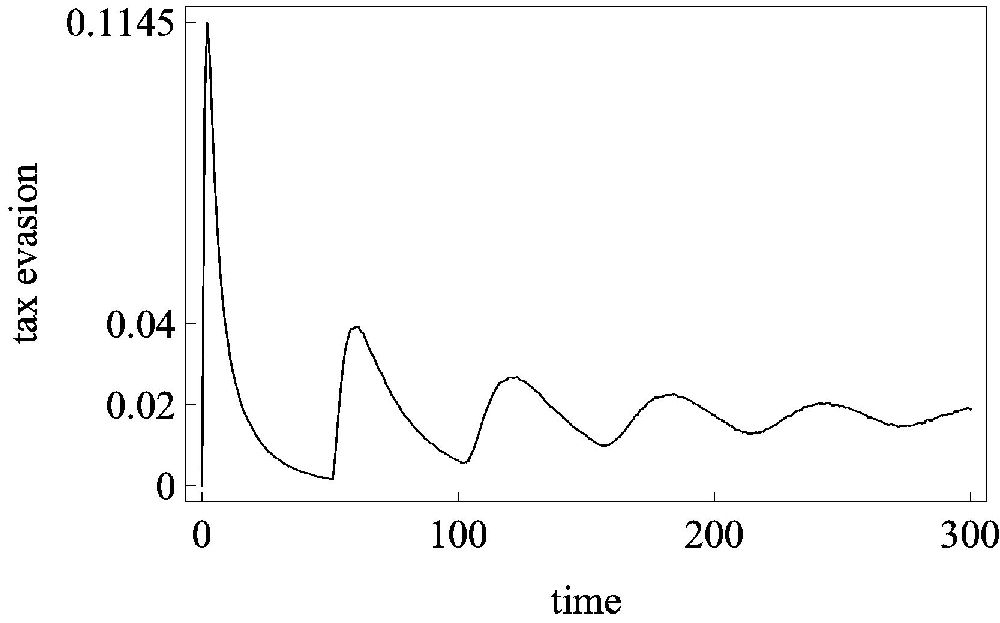}

\hspace{0.8cm} 

\includegraphics[width=7.0cm]{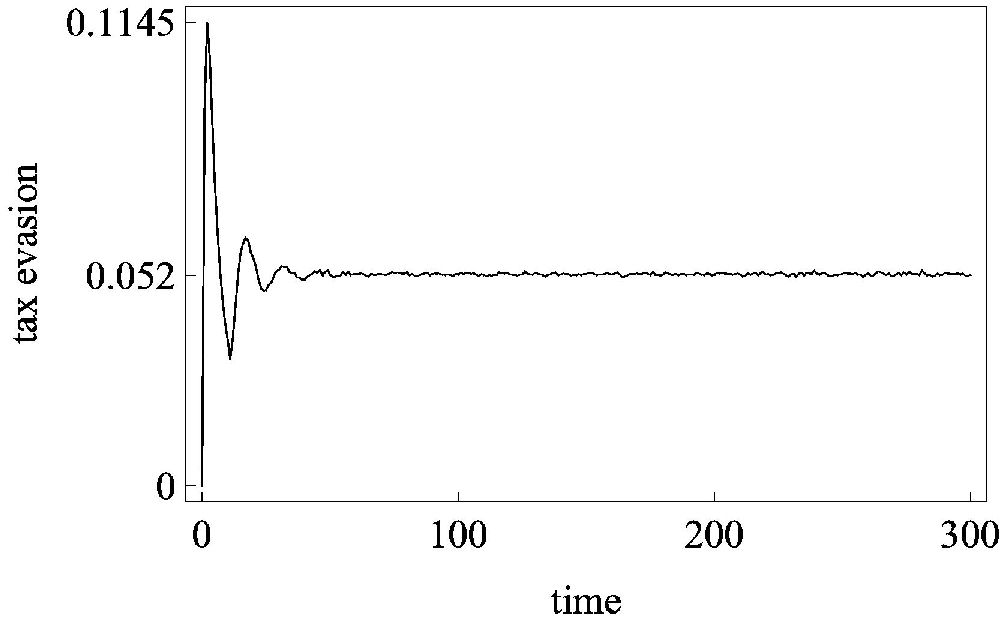}\\[0.8cm]

\noindent \hspace{0.0cm}\textbf{$\;\;\;k=50$, $p_a=0.05$} \hspace{3.9cm}  \textbf{$\;\;\;\;\;\;\;\; k=10$, $p_a=0.05$}\\[0.0cm]

\hspace{-0.4cm}

\includegraphics[width=7.0cm]{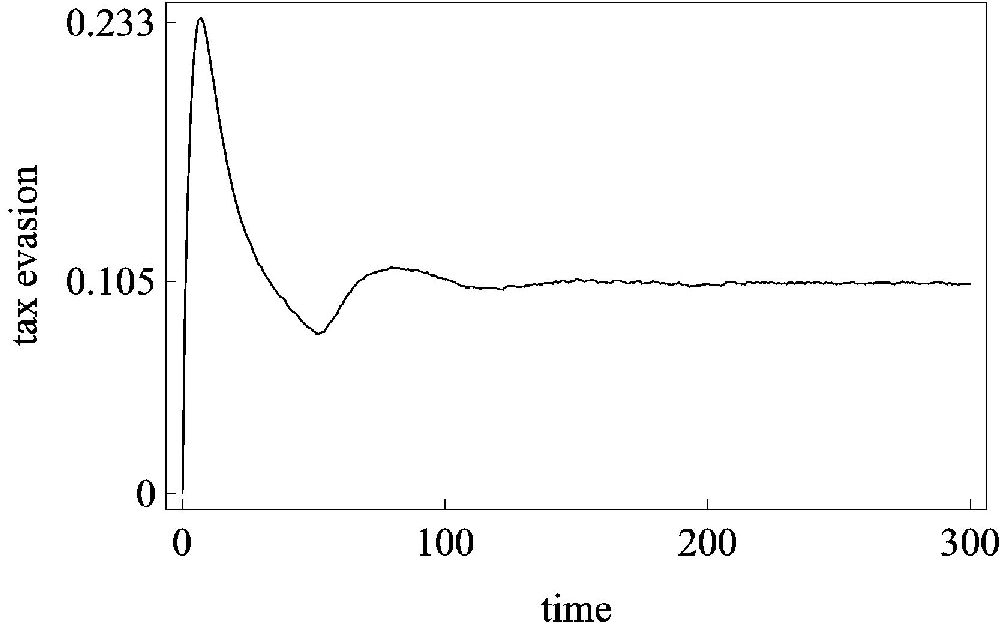}\hspace{0.8cm} 

\includegraphics[width=7.0cm]{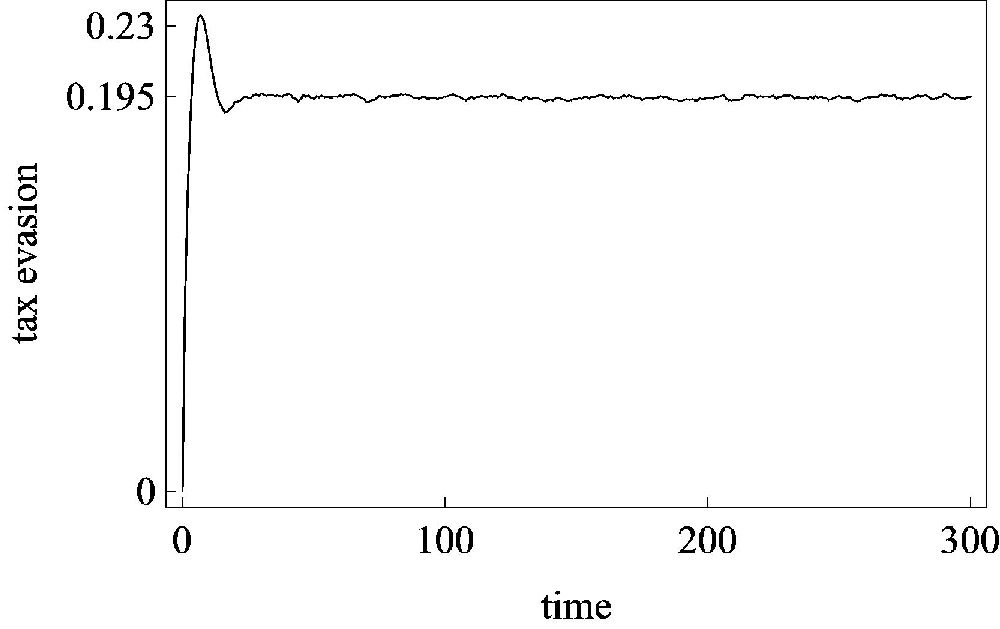}

\end{tabular}

\end{center}

\vspace{0.8cm}

\begin{center}

{\footnotesize{\noindent Figure 2: Tax evasion dynamics when the 

temperature is held constant at 3.\\ The same simulation design as in figure 1.}}

\end{center}


\newpage

 \vspace*{0.0cm}

\begin{center} 

\textbf{Temperature: 2.5}\\[1.0cm]  

\begin{tabular}{c}

\hspace{-1.9cm}\textbf{$\;\;\;\;\;\qquad\qquad k=50$} \hspace{6.5cm} \textbf{$k=10$}\\[0.0cm]

\includegraphics[width=7.8cm]{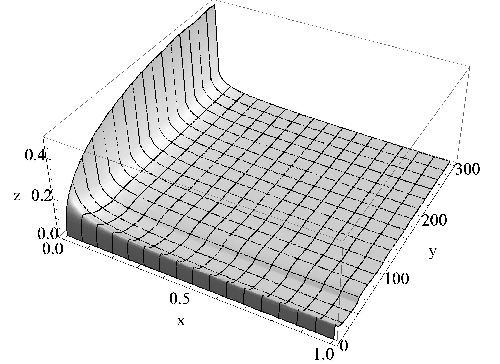}\hspace{-0.0cm}

\includegraphics[width=7.8cm]{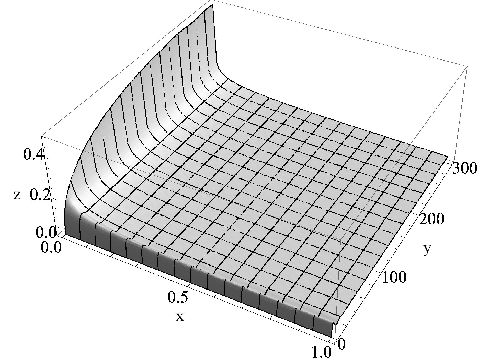} \\[0.8cm]

\noindent \hspace{0.0cm}\textbf{$\;\;\;\;k=50$, $p_a=0.9$} \hspace{3.9cm}  \textbf{$\;\;\;\;\;\;\;\; k=10$, $p_a=0.9$}\\[0.0cm]

\hspace{-0.4cm}

\includegraphics[width=7.0cm]{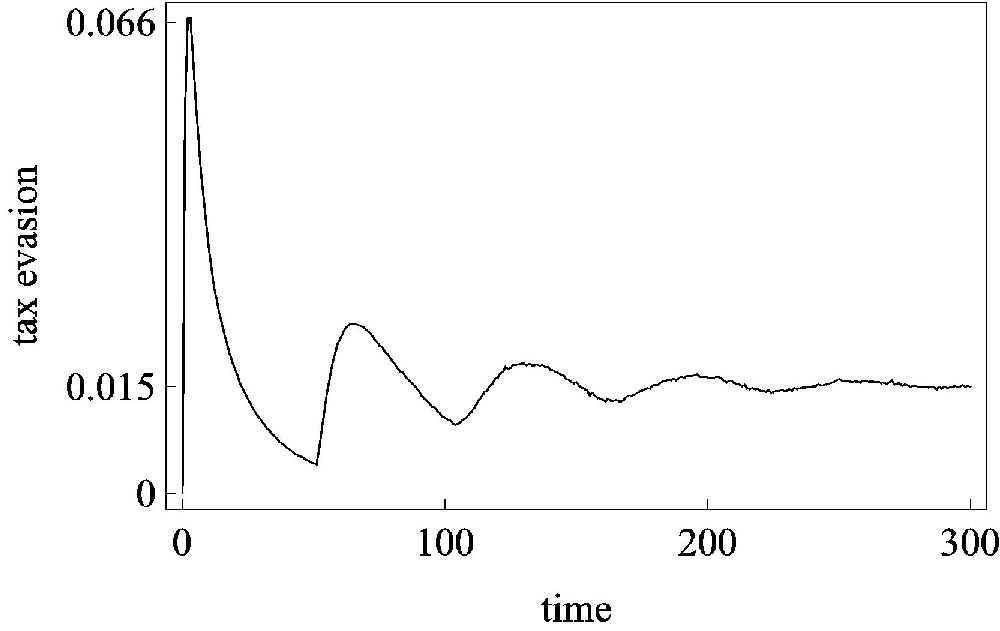}

\hspace{0.8cm} 

\includegraphics[width=7.0cm]{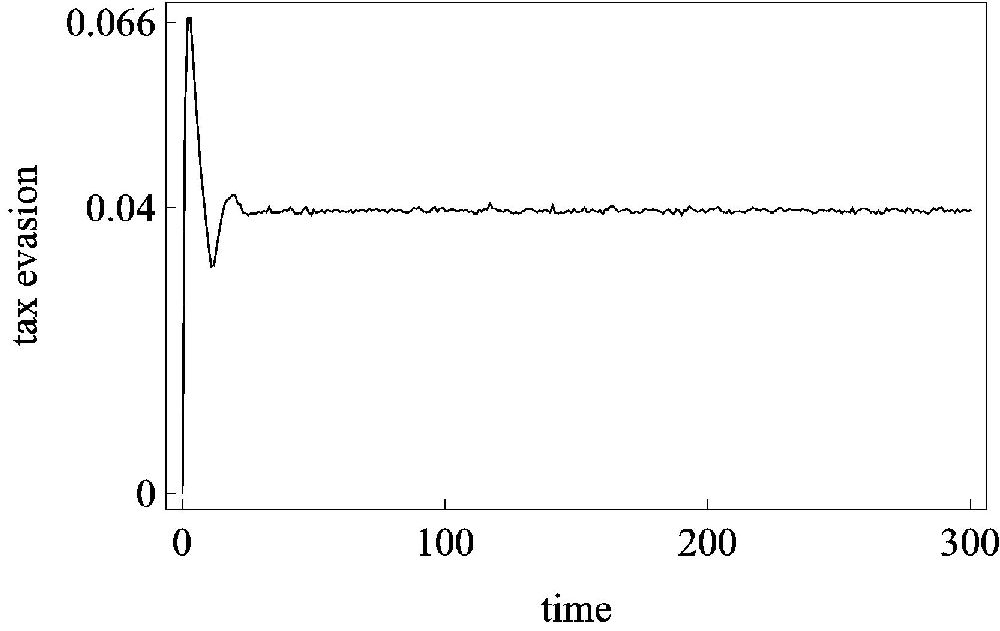} \\[0.8cm]

\noindent \hspace{0.0cm}\textbf{$\;\;\;k=50$, $p_a=0.05$} \hspace{3.9cm}  \textbf{$\;\;\;\;\;\;\;\; k=10$, $p_a=0.05$}\\[0.0cm]

\hspace{-0.4cm}

\includegraphics[width=7.0cm]{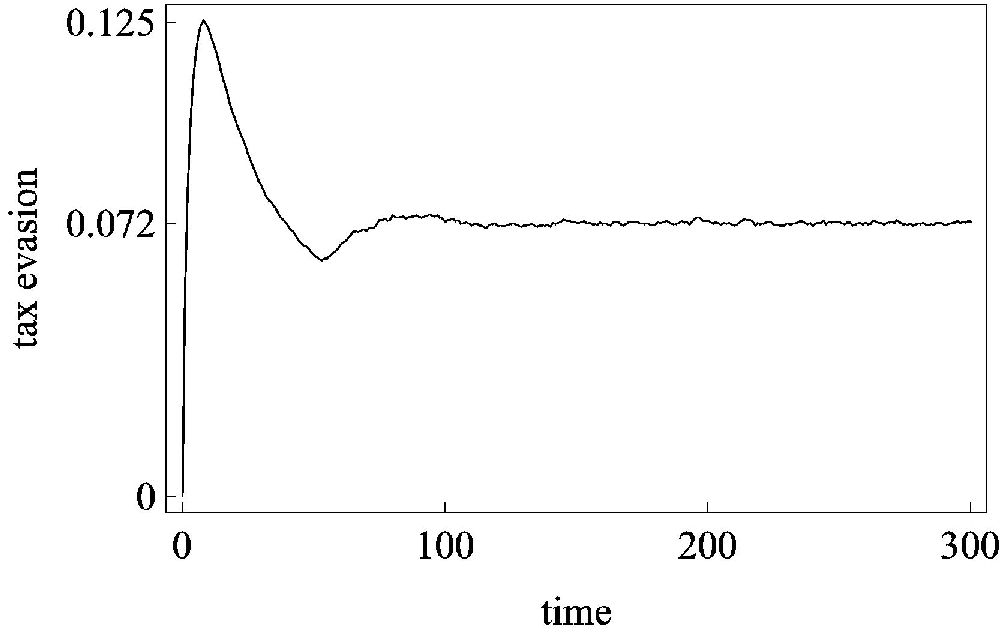}\hspace{0.8cm} 

\includegraphics[width=7.0cm]{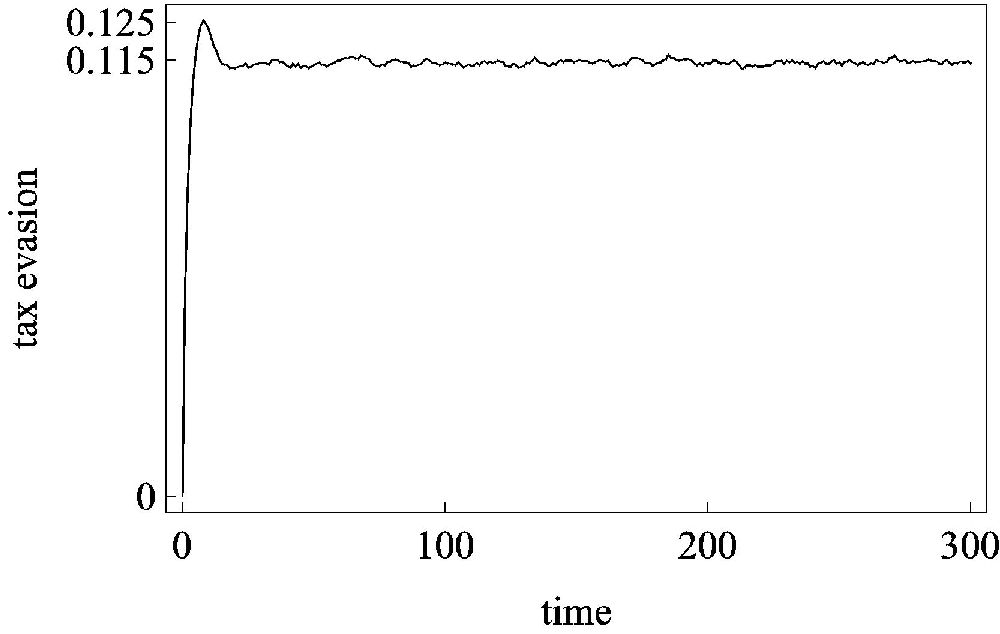}

\end{tabular}

\end{center}

\vspace{0.8cm}

\begin{center}

{\footnotesize{\noindent Figure 3: Tax evasion dynamics when the 

temperature is held constant at 2.5.\\ The same simulation design as in figure 1.}}

\end{center}


\newpage

 \vspace*{0.0cm}

\begin{center}  

 \textbf{Temperature: 2.0}\\[1.0cm] 

\begin{tabular}{c}

\hspace{-1.9cm}\textbf{$\;\;\;\;\;\qquad\qquad k=50$} \hspace{6.5cm} \textbf{$k=10$}\\[0.0cm]

\includegraphics[width=7.8cm]{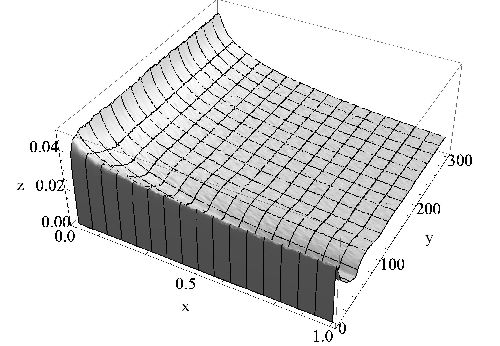}\hspace{-0.0cm}

\includegraphics[width=7.7cm]{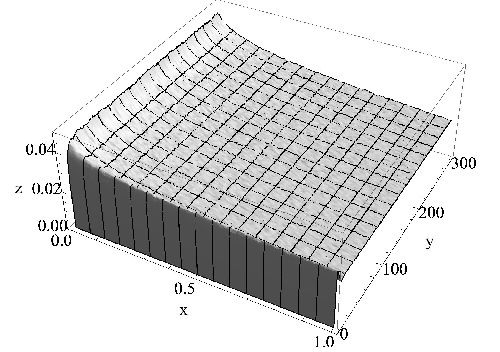} \\[0.8cm]

\noindent \hspace{0.0cm}\textbf{$\;\;\;\;k=50$, $p_a=0.9$} \hspace{3.9cm}  \textbf{$\;\;\;\;\;\;\;\; k=10$, $p_a=0.9$}\\[0.0cm]

\hspace{-0.4cm}

\includegraphics[width=7.0cm]{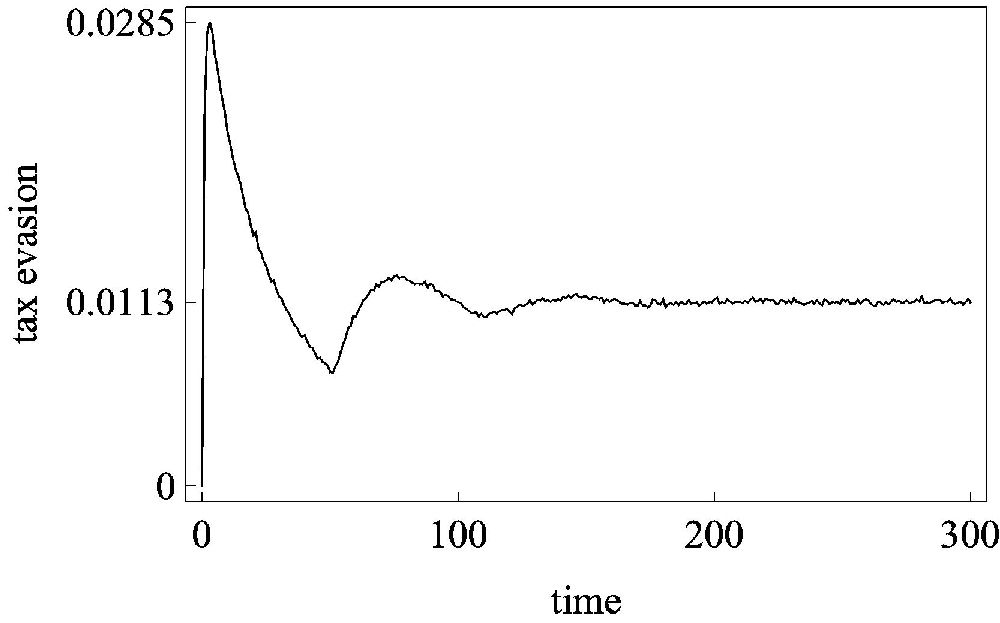}

\hspace{0.8cm} 

\includegraphics[width=7.0cm]{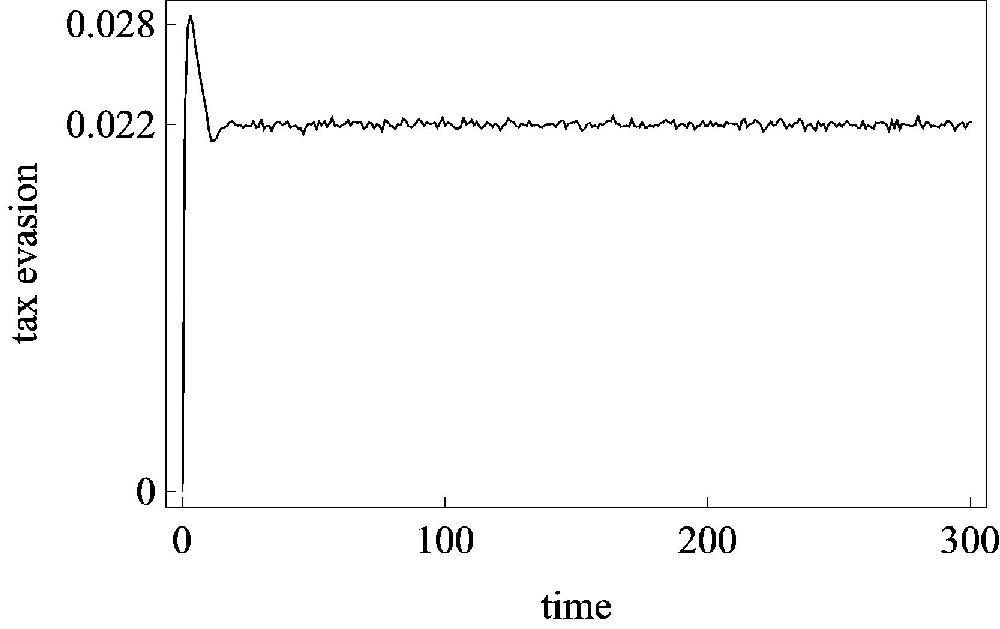} \\[0.8cm]

\noindent \hspace{0.0cm}\textbf{$\;\;\;k=50$, $p_a=0.05$} \hspace{3.9cm}  \textbf{$\;\;\;\;\;\;\;\; k=10$, $p_a=0.05$}\\[0.0cm]

\hspace{-0.4cm}

\includegraphics[width=7.0cm]{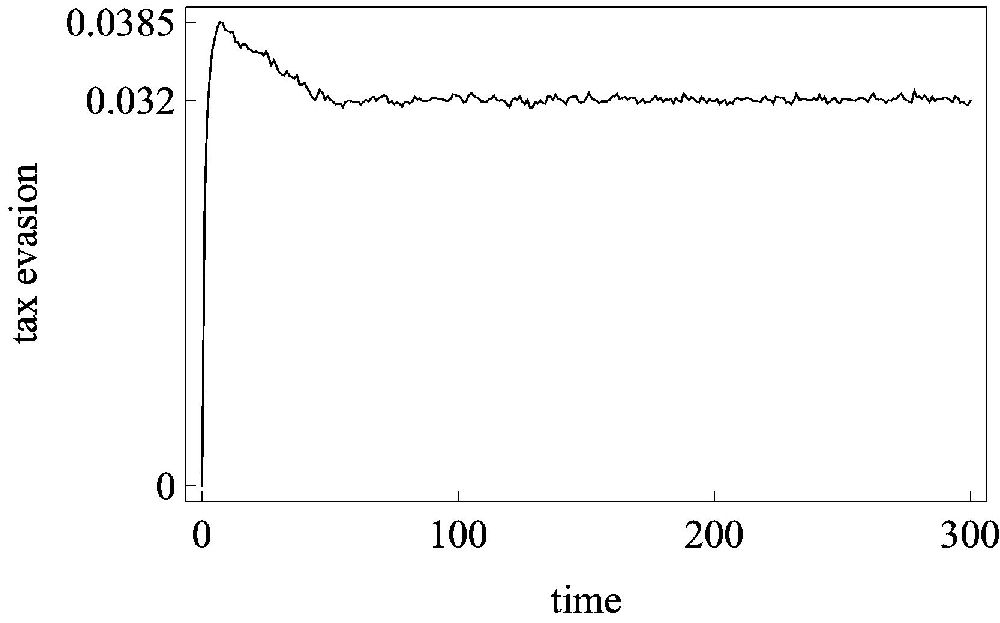}\hspace{0.8cm} 

\includegraphics[width=7.0cm]{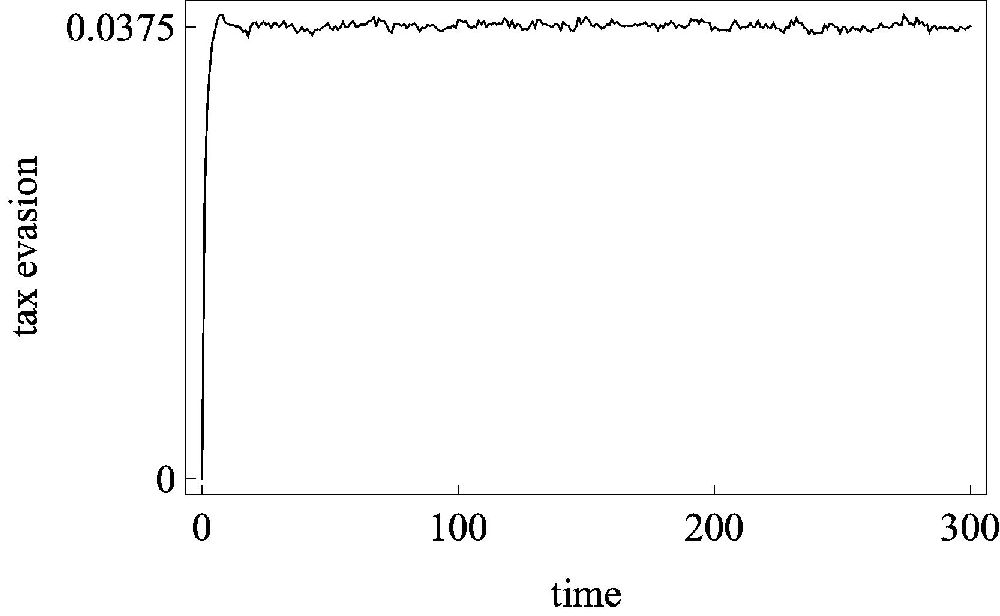}

\end{tabular}

\end{center}

\begin{center}

\vspace{0.8cm}

{\footnotesize{\noindent Figure 4: Tax evasion dynamics when the 

temperature is held constant at 2.\\ The same simulation design as in figure 1.}}

\end{center}

\end{document}